\documentclass[fleqn,usenatbib]{mnras}

\usepackage{newtxtext,newtxmath}

\usepackage[T1]{fontenc}
\usepackage{ae,aecompl}

\usepackage{graphicx}	
\usepackage{amsmath}	
\usepackage{amssymb}	

\renewcommand{\ion}[2]{#1\,{\sc #2}}
\newcommand{\cm}{cm$^{-3}$}

\title{Quiet Sun electron densities and their uncertainties derived from spectral emission line intensities}

\author[K. P. Dere]{
Kenneth P. Dere \thanks{E-mail: kdere@gmu.edu}
\\
Department of Physics and Astronomy, George Mason University, 4400 University Dr, Fairfax, VA, 20023, USA\\
}

\date{Accepted XXX. Received YYY; in original form ZZZ}

\pubyear{2020}

\begin{document}

\label{firstpage}
\pagerange{\pageref{firstpage}--\pageref{lastpage}}
\maketitle

\begin{abstract}
The goal of this paper is to apply statistical methods to determine electrons densities and their errors from measurements of density-sensitive line intensities in the quiet Sun. Three methods are employed. The first is the use of L-function plots to provide a quick visual assessment of the likelihood that a set of line intensities can provide a robust estimate of these quantities. A second methods involves a $\chi^2$ minimization together with a prescription for determining the regions of statistical confidence in addition to the best-fitting value. A third method uses a Bayesian inference technique that employs a Monte-Carlo Markov-chain (MCMC) calculation from which an analysis of the posterior distributions provide estimates of the mean and regions of high probability density. Using these three methods, observations of extreme-ultraviolet spectral lines originating from regions of the quiet Sun have been analyzed. The quantitative $\chi^2$ minimization and MCMC sampling provide results that are generally in good agreement, especially for sets of lines of ions that have L-function plots that suggest that a robust analysis might be possible.
\end{abstract}




\section{Introduction}
\label{sec:Intro}

The physical state of an astrophysical plasma is defined by several parameters including the electron density. Perhaps the first use of density sensitive line ratios to determine electron densities was that by \citet{seaton_1954}, who analyzed the spectrum of several nebulae. Since that time, there have been many improvements in our ability to calculate the atomic parameters that are needed to provide the  theoretical ratios as functions of density and temperature. In particular, the development of the  atomic structure code \textsc{SUPERSTRUCTURE} \citep{superstructure}, the distorted wave (DW) electron scattering code of \citet{eissner_dw}, the atomic code of \citet{cowan} and others played a strong part in the ability to understand coronal emissions. For example, based on the calculations of \citet{blaha_fe14}, \citet{purcell_widing} determined the electron density in a solar flare from the lines of \ion{Fe}{xiv}. \citet{cowan_widing} derived electron densities of a solar flare from \ion{Fe}{xv} spectra. \citet{flower_nussbaumer_fe13} determined electron densities from ratios of \ion{Fe}{xiii} lines measured by \citet{malinovsky_heroux} in an active corona. In the intervening half century, there have been many improvements in spectral codes to that allow more accurate interpretations of the spectra. Further, there have been numerous experiments with improved spatial and spectral resolution and sensitivity. Many of these have been reviewed by \citet{delzanna_lr}. The point of this paper is not concerned so much with determining electron densities from observed spectra but to point out techniques that allow the determination of the degree of statistical confidence that one can place in these diagnostic results.

In this analysis, the observed EUV line intensities in the quiet Sun from the Solar EUV Rocket and Spectrograph (SERTS) rocket flight in 1993 \citep{brosius} are used. This data set contains a number of density-sensitive line ratios of several ions. Three methods are used to arrive at {\it best values} for the electron densities and their statistical uncertainties. The first method is a visualization of the situation that allows the user to develop a reasonably good estimate of the electron density and its range of uncertainty and that of the emission measure. In some cases, it will be clear that there is no {\it best value} and perhaps only a lower limit or an upper limit to the density can be determined. This visualization is displayed by plotting a simplified version of the L-functions of \citet{landi-L-function}. The L-function of each line is the observed intensity of the line divided by the contribution function as a function of electron density. The first such plot can be seen in Fig.~\ref{fig:fe_13_tab2_emplot}. It should be noted that the method of determining the contribution function has been simplified from that of \citet{landi-L-function}. This will be further discussed in Section~\ref{subsec:L-functions}.

The second method of analysis, uses a $\chi^2$ minimization technique described in Section~\ref{subsec:brute}. The estimate of the emission measure gathered from the L-functions serves as a starting point for the minimization. The value of $\chi^2$ is then found by an iteration over a range the electron density from 10$^6$ to 10$^{12}$ cm$^{-3}$.

The third method of analysis is through Bayesian inference by means of a Monte-Carlo Markov Chain (MCMC). Here, the \textsc{PyMC} \citep{pymc} \textsc{Python} package is used. This is discussed in Section~\ref{subsec:mcmc}.

\section{Observations}
\label{sec:Observations}

\citet{brosius} reported spectral line intensities obtained with the SERTS from two flights in 1991 and 1993. Spectral line observations were made in a wavelength range between 274 and 417 \AA. The spectral lines are formed over a temperature range from about 10$^5$ to 2 $\times$ 10$^6$ K, \citep{brosius}. The intensities tabulated by \citet{brosius} were obtained by averaging over the 282 arcsec slit. The quoted spatial resolution is about 5 arcsec. Consequently, the effects of any substructures are likely to be washed out and any densities derived from these spectra will be average densities. The data set includes line intensity ratios that are functions of electron density and these have been previously analyzed by \citet{brosius}. In this paper I will report the analysis of the 1993 quiet Sun spectra. The 1991 quiet Sun spectra have also been analyzed to check the consistency between the two data sets. For the sake of brevity, only the analysis of the 1993 quiet Sun data will be reported here. 

Density-dependent emission line ratios of \ion{Mg}{viii}, \ion{Si}{ix}, \ion{Si}{x}, \ion{Fe}{xi}, \ion{Fe}{xii}, \ion{Fe}{xiii}, and \ion{Fe}{xiv} are available in this set of observations and will be examined here.

\section{Uncertainties}
\label{sec:uncertainties}

In order to determine the uncertainties in the derived electron densities, it is necessary to understand the ability to reproduce the observed intensities. The uncertainties in the electron densities derived from observed spectra come from two different sources. The first is the errors in the observed spectral line intensities and the second is the errors in the atomic models. Here, I will only compare lines of the same ion so that the uncertainties in the atomic model lie only in the excitation and radiative rates. It is also necessary to specify a temperature for each ion so there is also some uncertainty in that decision. In this analysis, it is not possible to separate the measurement errors from any errors in the atomic models.

\subsection{Uncertainties in the measured line intensities}
\label{subsec:errors}

\citet{brosius} provide estimates of the intensity uncertainties for each of the observed lines. An analysis of the errors of the intensities reported by \citet{brosius} in their Table 2 indicates that the most probable ratio of the uncertainty of the intensity is about 0.12. The values range from a minimum of 0.11 to a maximum of 0.5. Three intensities have a relative error lying outside 3$\sigma$.

A brief inspection of Table 6 of \citet{brosius} suggests the the uncertainties of the observed density- and temperature-insensitive line ratios of the same ion are of the order of 15-25 per cent. The authors also state that the relative calibration is accurate to about 20 per-cent. As this source of error mainly applies when comparing lines at widely separated wavelengths, this is probably included in the 15-25 per-cent error just quoted.  The analyses performed below have been repeated numerous times in order to determine the uncertainties to use in estimating the weighted chi-square function. These uncertainties are generally about 20 per cent.

\subsection{Uncertainties in the atomic models}
\label{subsec:model_uncertainties}

It is difficult to estimate {\it a priori} the errors that may exist in the atomic data used to construct the atomic models. Recently, \citet{yu} have considered the effect of uncertainties in the atomic parameters on the emissivities and emissivity ratios as a function of electron density for the case of \ion{Fe}{xiii} that is also considered here in Section~\ref{subsec:fe_13}. They have simulated the effect of errors in the collision rates and A-values and calculated the emissivities of several \ion{Fe}{xiii} lines near 200 \AA. Based on a comparison between 2 sets of atomic calculations, they assign errors of 5 per cent for strong transitions, 10 per cent for weaker transitions and 30 per cent for the weakest transitions.

\citet{yu} further performed two Bayesian simulations to estimate the densities derived from the set of observed \ion{Fe}{xiii} line intensities from the prior set of simulated emissivities. Their estimates for the errors in the derived electron densities range from about 2 per cent to 10 per cent, depending on the method used.

Recently \citet{delzanna_n_4} examined the effect of using different atomic calculations to interpret line intensities of \ion{N}{iv}. They found that the difference in atomic calculations increases when considering weak transitions or between widely separated levels but that for the important \ion{N}{iv} lines, the various atomic models agree to about 20 per cent.

There have been a number of laboratory observations of lines emitted by the ions examined here and the measured intensities of the spectral lines compared with the intensities predicted by theoretical calculations. \citet{weller_fe_13_14_lab} measured line intensities of \ion{Fe}{xiii} through \ion{Fe}{xvi} from plasma discharges at the National Spherical Torus-Upgrade. They were able to independently determine the electron density from laser Thomson scattering. The electron densities in the Torus were on the order of 10$^{13}$ cm$^{-3}$. At these densities, the ratios of the spectral lines measured by \citet{weller_fe_13_14_lab} are at their high density limit.  They conclude that the measured intensity ratios agree with ratios calculated with the CHIANTI database (version 8.0.2 \citet{delzanna_v8}) to within about 30 per cent.

\citet{beiersdorfer_fe12_13_14} used an electron beam ion trap (EBIT) for the ions \ion{Fe}{xii, xiii, xiv}. Their measurements were performed at electron densities ranging from 2 $\times$ 10$^{10}$ to 2 $\times$ 10$^{11}$ cm$^{-3}$. The measured intensities were compared with predictions made with the \textsc{Flexible Atomic Code} (FAC), \citep{fac}. For the density-insensitive ratios of \ion{Fe}{xii}, they find good agreement between the measured and the predicted ratios. For one of the \ion{Fe}{xii} density-sensitive ratios a good agreement between measured and predicted ratios was found but for another, the difference was about 50 per cent. For their comparisons with \ion{Fe}{xiii} one of the density insensitive ratios had a measured value about 70 per cent higher than predicted. For several density-sensitive ratios, the agreement was on the order of 20 per cent. For \ion{Fe}{xiv} they found reasonable agreement.

\subsection{Uncertainties used in the analysis}
\label{subsec:uncertainties_used}

Based on the discussions in the previous two sections, it appears that it would be appropriate to consider that the error in reproducing the observed line intensities is about 20 per cent and this value has been used in the current analysis.

\section{Spectral line intensities}

\subsection{Calculation of spectral line intensities}
\label{subsec:calculation}

A method for the analysis of spectral line intensities for allowed lines was presented by \citet{pottasch} and has been used in similar ways up to the present:

\begin{equation}
    I = \int\, G(n_e, \,T) \, n_e \, n_H \, ds
\end{equation}

where the intensity I is in erg cm$^{-2}$ sr$^{-1}$ s$^{-1}$, n$_e$ is the electron density, T is the temperature, n$_H$ is the hydrogen density, both in cm$^{-3}$. The integral is performed along the line of sight. {\it G} is called the {\it Contribution Function}. The Contribution Function includes the emissivities of the line as well as the elemental abundance relative to hydrogen and the ionization equilibrium appropriate to the ion. When comparing lines from the same ion, the elemental abundance is not relevant and the ionization equilibrium only comes into account when specifying the temperature. 

For ions containing metastable levels, the calculation of the Contribution Function becomes more complicated and it is necessary to have a model of the ion including excitation and decay rates for a number of levels.

The model ions are provided by the CHIANTI atomic database \citep{dere_v1, dere_v9}. In addition to excitation and decay rates, the CHIANTI database provides tables of solar abundances from published sources and calculations of ionization equilibria. For this work, the abundances of \citet{scott_abund} and the ionization equilibrium calculated from the CHIANTI ionization and recombination rates \citep{dere_v6, landi_v7.1, dere_v9} are used.

To reproduce the intensities of a set of lines from a single ion it only necessary to know the temperature, the electron density and the emission measure. The contribution function for each ion has been determined by matching all lines in the CHIANTI database for elements with an abundance greater than 10$^{-7}$ the abundance of hydrogen that produce a spectral line within 0.05 \AA\ of the observed wavelength. This allows any problems with line blends to be evaluated. The lines used in this analysis are fairly strong and blending has not proven to be a significant problem. The contribution function is calculated for 241 density values ranging from 10$^6$ to 10$^{12}$ cm$^{-3}$.  This is performed with the ChiantiPy \citep{landi_v7.1, dere_v9} Python software package.

For all three methods outlined next in Section~\ref{sec:Methods}, the same values for the contribution functions have been used.

\section{Methods}
\label{sec:Methods}

\subsection{Simplified L-functions}
\label{subsec:L-functions}

To determine the electron density and its uncertainty from observations of spectral lines, it is best to consider all of the observed lines of a given ion at the same time. In the past, it has been typical to consider the theoretical ratio of two density-sensitive spectral lines and compare that ratio to the observed ratio. The uncertainties in the derived ratio are then determined from a knowledge of the uncertainties in the individual observed ratios. Using all of the lines of a given ion, an estimate of the {\it best value} is found by considering the L-functions. The L-function is the spectral line intensity divided by the contribution function G(n$_e$, T) as a function of electron density n$_e$ at a specific temperature T. \citet{landi-L-function} used the average of the temperature weighted by the emission measure and the contribution function (their Eq. 16). In this analysis, I have used the temperature that  maximizes the contribution function. In Section \ref{subsubsec:temperature-dependence} the dependence on the temperature is examined in the case of Fe XIII. The L-functions used here will be referred to as {\it simplified}  L-functions in order to clarify the difference. Examples of L-functions for a variety of ions are provided by \citet{landi-L-function} and here, in Section~\ref{subsec:fe_13} and afterwards.

From the simplified L-functions displayed in Fig.~\ref{fig:fe_13_tab2_emplot}, a visual inspection suggests that the best guess for the electron density would be on the order of 10$^9$ cm$^{-3}$ and 10$^{27}$ cm$^{-5}$ for the line-of-sight emission measure ($\int n_e n_H ds$).

\subsection{Chi-squared minimization}
\label{subsec:brute}

The minimization of $\chi^2$ has been used as a technique for solving non-linear problems for some time. 
\begin{equation}
    \chi^2 = \sum_i ((I_i - P_i)/\sigma_i)^2
\end{equation}
where I$_i$ is the observed intensity for line i, P$_i$ is the predicted intensity and $\sigma_i$ is an estimate of the combined error in observed and predicted intensities. A value of $\sigma$ = 0.2*I$_i$, is used, following Section~\ref{subsec:errors}. 

A key point of this paper is to determine the uncertainties in the derived densities and \citet{lampton} have provided a prescription for determining the range of confidence in the $\chi^2$ minimized solution. See Equation 5 of \citet{lampton}.
\begin{equation}
    \chi^2 \le \chi^2_{min} + \chi^2_p(\alpha)
        \label{eq:lampton}
\end{equation}
where $\chi^2$ is the value of $\chi^2$ in some region of parameter space, $\chi^2_{min}$ is the minimum value and $\chi^2_p(\alpha)$ is the value of the $\chi^2$ function for p degrees of freedom and significance $\alpha$. These values are commonly tabulated and can be calculated with the \textsc{Scipy Python} package. For a given level of confidence $\alpha$ and degrees of freedom {\it p} the regions of confidence is where $\chi^2$ is less than the right hand side of the equation. The number of degrees of freedom is 2, electron density and emission measure. For the present  analysis, values of the significance of 0.32 and 0.05 are used, corresponding to a level of confidence of 0.68 and 0.95 and roughly to  1$\sigma$ and 2$\sigma$ errors.

Recently, \citet{lucy_2016} has determined a somewhat simpler criterion for cases where the model is linear in its parameters and the measurement errors obey a normal distribution. 
\begin{equation}
     \chi^2 \le \chi^2_{min} + p
        \label{eq:lucy}
\end{equation}
It is not clear that the conditions for use of this prescription are always met in the present analysis. Consequently, the prescription of \citet{lampton} will be used here.

For each of the 241 values of electron density, the value of the best-fitting emission measure is calculated to determine the value of $\chi^2$. The solution is provided by the \textsc{Scipy leastsq} function that employs a Levenberg-Marquardt algorithm. The initial value of the search for the optimal value of the emission measure begins with the value that is estimated from the plot of the simplified L-functions. A plot of $\chi^2$ {\it vs} electron density can be seen in Fig~\ref{fig:fe_13_tab2_chisquared}. There, the values of $\chi^2$ corresponding to levels of confidence of 0.68 and 0.95 are also plotted and from these the regions of confidence are determined.

\subsection{Bayesian inference by means of a Monte-Carlo Markov-chain technique}
\label{subsec:mcmc}

The use of Bayesian inference in the analysis of astrophysical data has become widespread as computers have become faster and packages for performing MCMC sampling are  made available. For the present analysis, the \textsc{PyMC Python} package of \citet{pymc} has been employed. A review of the use of MCMC techniques for Bayesian inference has been given by \citet{sharma_bayes}. By creating a model as a function of one or more parameters to reproduce the observed data, one then uses an MCMC approach to produce the posterior distributions based on a set of prior assumptions. The Metropolis-Hastings step method is the \textsc{PyMC} default and has been used here.

As discussed in Section~\ref{subsec:L-functions}, the model consists of set of contribution functions for each spectral line as a function of electron density, at a single temperature. From the plots of the simplified L-functions, it is relatively straightforward to estimate values of the density and emission measure that will give a good reproduction of the observed intensities. The prior distribution for the emission measure is a Normal distribution centered on the value deduced from the simplified L-functions. The prior distribution for the index of the electron density is an uninformed discrete uniform distribution. The use of the index allows for quick sampling of the pre-calculated density grid of the contribution function. The value of the prior for the density index is set to the value expected from the simplified L-function plot. After a burn-in period, the model samples the density space and after a specified number of iterations, the posterior distributions for the density index and the emission measure are available.

\section{Analysis}
\label{sec:analysis}

\subsection{Analysis of the emission lines of Fe XIII}
\label{subsec:fe_13}

EUV spectral lines of \ion{Fe}{xiii} have been used in the past for the analysis of electron densities in the solar corona \citep{flower_nussbaumer_fe13}. The ion \ion{Fe}{xiii}, out of the set of observed ions, is the first ion selected for presentation as it provides the most reliable assessment of electron densities in the quiet Sun when used with the lines found in Table 2 of \cite{brosius}. The \ion{Fe}{xiii} lines used here are listed in Table~\ref{tab:tab2_fe_13_linelist}.

\begin{table}
	\begin{center}
	\caption{\ion{Fe}{xiii} line list, T = 1.78 $\times$ 10$^6$ K}
	\label{tab:tab2_fe_13_linelist}
	\begin{tabular}{ccrrr} 
		\hline
		$\lambda_B$ (\AA) & $\lambda_C$  (\AA)&  I$_{obs}$ & I$_{best}$ & I$_{pymc}$\\
		\hline
		 311.574 &   311.547  & 6.13 & 2.9 & 2.8 \\
		312.171 &  312.174 &   17.80 & 17.6 & 17.6 \\
        312.907 &  312.868 &  7.34 &  6.2 & 6.0 \\
        318.129 &  318.130 &  6.09 &  8.9 & 8.7 \\
        320.802 &   320.800 &     24.50 & 20.4 & 20.0 \\
        321.464 &   321.466 &  8.64 &  8.8 & 8.8 \\
        348.196 &    348.183 &    55.00 & 43.1 & 43.5 \\
        359.658 &    359.644 &   28.40 &  29.4  & 28.7 \\
        359.851 &    359.839 &   11.80 & 11.3 & 11.4 \\
        \hline
	\end{tabular}
	\end{center}
	
	\raggedright{$\lambda_B$ is the wavelength measured by Brosius, $\lambda_C$ is the wavelength from CHIANTI, I$_{obs}$ is the observed SERTS intensity (erg cm$^{-2}$ sr$^{-1}$ s$^{-1}$), I$_{best}$ is the best-fitting intensity from the $\chi^2$ minimization, I$_{pymc}$ is the mean intensity from the \textsc{PyMC} sampling.}
\end{table}

\begin{figure}
	\includegraphics[width=\columnwidth]{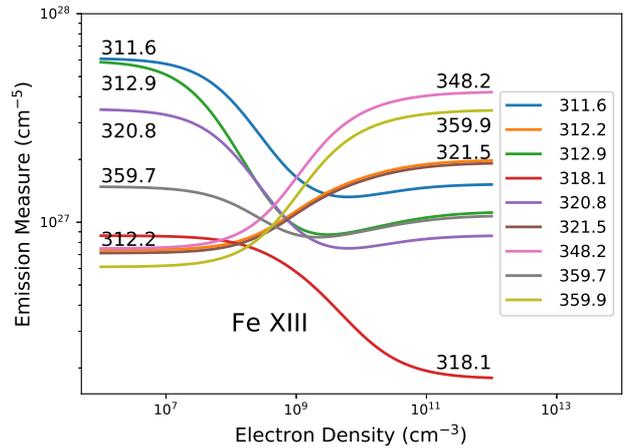}
    \caption{The simplified L-functions of the \ion{Fe}{xiii} lines. It should be noted that the lines at 312.2 and 321.5 \AA\ have essentially overlapping L-functions.}
    \label{fig:fe_13_tab2_emplot}
\end{figure}

From the simplified L-functions displayed in Fig.~\ref{fig:fe_13_tab2_emplot}, a visual inspection suggests that the best estimate for the electron density would be on the order of 10$^9$ cm$^{-3}$ and 10$^{27}$ cm$^{-5}$ for the line-of-sight emission measure ($\int n_e n_H ds$). Fig.~\ref{fig:fe_13_tab2_emplot} also suggests that it should be possible to determine the regions of certainty for the electron densities derived from the \ion{Fe}{xiii} lines.

\subsubsection{Density determination by Chi-Squared minimization}

\begin{figure}
	\includegraphics[width=\columnwidth]{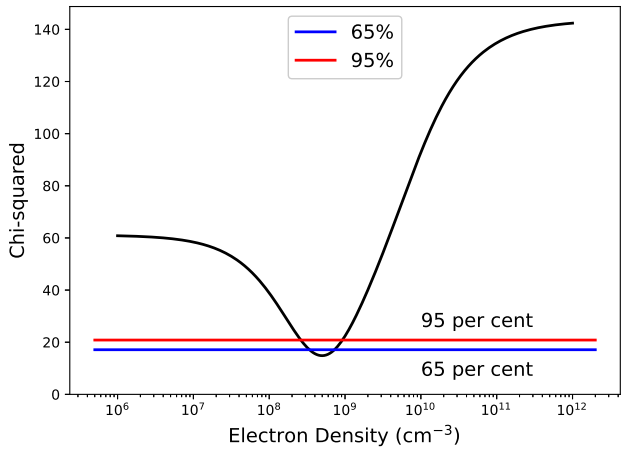}
    \caption{$\chi^2$ for the \ion{Fe}{xiii} lines as a function of density together with the levels of 68 per cent and 95 per cent confidence.}
    \label{fig:fe_13_tab2_chisquared}
\end{figure}

\begin{figure}
	\includegraphics[width=\columnwidth]{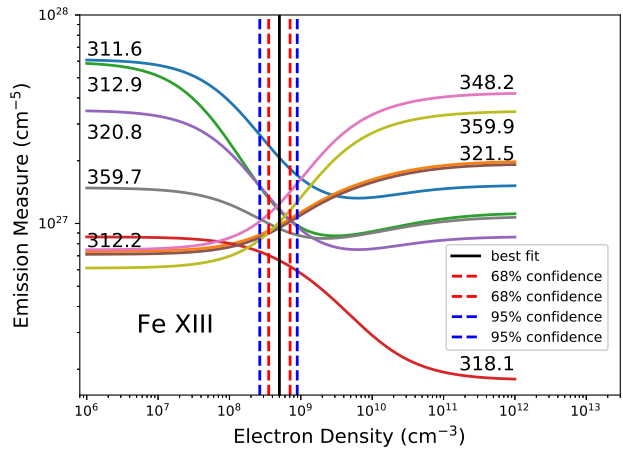}
    \caption{The emission measure needed to reproduce the intensities of the \ion{Fe}{xiii} lines as a function of electron density together with the best-fitting density and the 68 per cent and 95 per cent confidence limits derived by the $\chi^2$ minimization.}.
    \label{fig:fe_13_tab2_emplot_best}
\end{figure}

Using the calculations shown in Fig.~\ref{fig:fe_13_tab2_emplot}, it is possible to perform a simple search over electron density to determine the values of the density and emission measure that minimizes $\chi^2$. The value of $\chi^2$ as a function of electron density is shown in Fig.~\ref{fig:fe_13_tab2_chisquared}. The minimum occurs at a density value of 5 $\times$ 10$^8$ cm$^{-3}$ at an emission measure of 1 $\times$ 10$^{27}$ cm$^{-5}$. Also shown in Fig.~\ref{fig:fe_13_tab2_chisquared} are the horizontal lines at the minimum of $\chi^2$ plus $\chi^2_p(\alpha)$, where p = 2 and $\alpha$ = 0.32 corresponding to a confidence value of 68 per cent and $\alpha$ = 0.05 corresponding to a confidence value of 95 per cent, following the prescription of \citet{lampton}. From these data, the value of the best-fitting density and the regions of 68 per cent and 95 per cent confidence are found and  are shown in Fig.~\ref{fig:fe_13_tab2_emplot_best} together with the simplified L-functions from Fig.~\ref{fig:fe_13_tab2_emplot} and listed in Table~\ref{tab:tab2_fe_13_summary}. The best-fitting density of 5 $\times$ 10$^{8}$ cm$^{-3}$ is comparable to the our initial guess.

\subsubsection{Density determination by means of MCMC sampling with PyMC}
\label{subsubsec:fe13_pymc}

The application of the \textsc{PyMC} package for inferring electron densities and their uncertainties has been outlined above in Section~\ref{subsec:mcmc}, including the model and the prior distributions for the electron density and the emission measure.

\begin{figure}
	\includegraphics[width=\columnwidth]{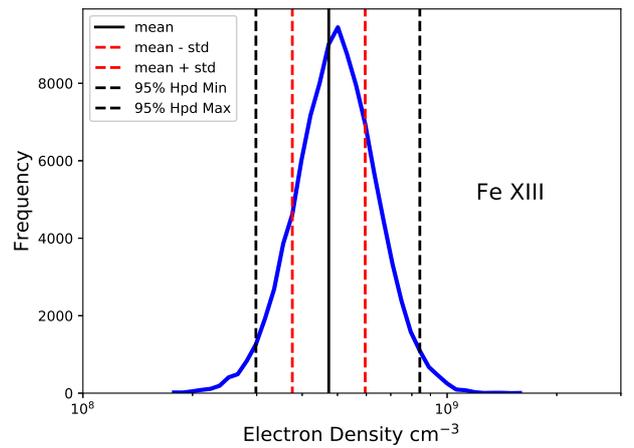}
    \caption{The posterior distribution of the electron density produced by the application of \textsc{PyMC} to the intensities of the \ion{Fe}{xiii} lines. Also plotted is the mean density,  the mean density $\pm$ the standard deviation (std) of the distribution and the region of 95 per cent high probability density (Hpd).}
    \label{fig:tab2_fe_13_density_hist}
\end{figure}

\begin{figure}
	\includegraphics[width=\columnwidth]{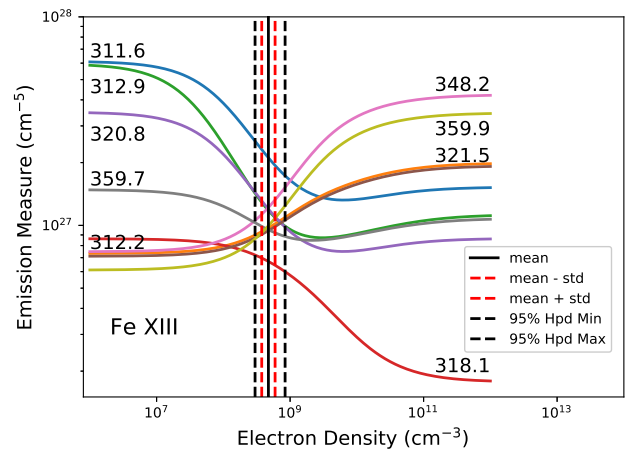}
    \caption{The emission measure needed to reproduce the intensities of the \ion{Fe}{xiii} lines as a function of electron density with the mean, standard deviation (std) and region of 95 per cent high probability density (Hpd)  derived from the \textsc{PyMC} sampling.}
    \label{fig:tab2_fe_13_emplot_w_mean_std_HPD}
\end{figure}

The histogram of the samples of electron density obtained from a run with \textsc{PyMC} are shown in Fig.~\ref{fig:tab2_fe_13_density_hist}. The \textsc{PyMC} software typically outputs the mean, the standard deviation and the limits of the 95 per cent high probability density region. These are also plotted in Fig.~\ref{fig:tab2_fe_13_density_hist} together with the simplified L-functions, are plotted in Fig.~\ref{fig:tab2_fe_13_emplot_w_mean_std_HPD} .

\subsubsection{Temperature dependence}
\label{subsubsec:temperature-dependence}

The choice of the temperature as that which maximizes the contribution function rather than the average over the differential emission measure as used by \citet{landi-L-function}  might affect the derived densities and errors. I have examined this by repeating that MCMC analyses with temperatures above and below the original choice of 1.78 $\times$ 10$^6$ K. The contribution function for the selected lines of  \ion{Fe}{xiii} fall to a factor of 3 below the maximum at temperatures of 1.4 $\times$ 10$^6$ and 2.2 $\times$ 10$^6$ K. The mean densities from these two analyses are 4 $\times$ 10$^8$ and 6 $\times$ 10$^8$ cm$^{-3}$, respectively. Thus it is possible that the derived electron densities could be off by about 20 per cent due to the choice of temperature. These values are at about the 68 per cent confidence limits and the standard deviation of the high probability density region. The results are tabulated in Table~\ref{tab:tab2_fe_13_summary} in Section~\ref{subsubsec:fe13_summary}.
It should be noted that these calculations provide an example of the possible effect on the choice to temperature on the resulting density estimate in the case of \ion{Fe}{xiii}. It is possible that the effect many differ in other ions.

\subsubsection{Summary of the analysis of Fe XIII}
\label{subsubsec:fe13_summary}

\begin{figure}
	\includegraphics[width=\columnwidth]{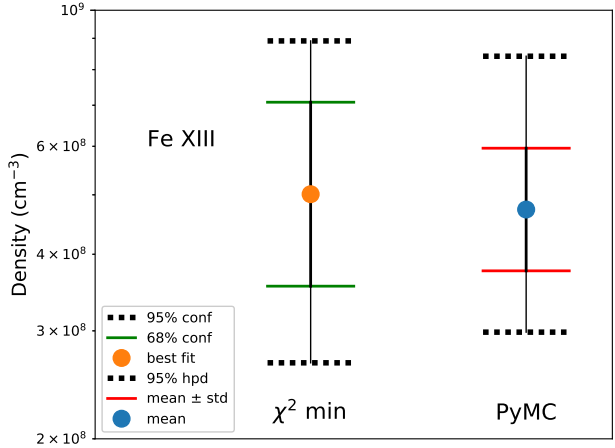}
    \caption{A comparison of the  \ion{Fe}{xiii} electron densities and their uncertainties derived by means of a $\chi^2$ minimization and by a MCMC sampling.}
    \label{fig:tab2_fe_13_br_mc_compare}
\end{figure}

A visual comparison of the results is shown in Fig.~\ref{fig:tab2_fe_13_br_mc_compare}. The numerical values derived from the $\chi^2$ minimization and the MCMC sampling are shown in Table~\ref{tab:tab2_fe_13_summary}. The electron density and their regions of confidence derived by the two methods are quite comparable.  

\begin{table*}
	\begin{center}
	\caption{Summary of the analyses of \ion{Fe}{xiii}}
	\label{tab:tab2_fe_13_summary}
	\begin{tabular}{ccccccc} 
		\hline
		  T(K) & 95\% Conf  & 68\% Conf  & Best-fit  & 68\% Conf  & 95\% Conf & RelStd(Int)\\
		\hline
      1.78e+6 & 2.7e+08 &    3.5e+08 &    5.0e+08 &    7.1e+08 &    8.9e+08 & 0.18 \\
        \hline
	  T (K) &	    95\% HPD  &  Std  &       Mean  &       Std  &   95\% HPD & \\
    \hline
     1.78e+6 &   3.0e+08 &    3.8e+08 &    4.7e+08 &    6.0e+08 &    8.4e+08 & 0.18 \\
    \hline
     1.4e+6 & 2.4e+8 &    3.e+8 &    4.e+08 &    5.e+08 &    7.e+8 & 0.19  \\ 
    \hline
      2.2e+6 &   3.5e+08 &    4.5e+08 &    5.6e+8 &    7.e+08 &    9.e+08 & 0.17 \\
   \hline
	\end{tabular}
	\end{center}
	\raggedright{Note:  95\% Conf refers to the lower and upper limits of the region of 95 per cent confidence, 68\% Conf refers to the lower and upper limits of the region of 68 per cent confidence, and Best-fit refers to the electron density that provides yields the minimum value of $\chi^2$, RelStd(Int) refers to the mean of the relative deviance of the observed intensities with respect to the predicted intensities, 95\% HPD denotes the lower and upper limits of the region of High Probability Density in the \textsc{PyMC} posterior distribution, Std denotes the electron density at the mean $\pm$ the standard deviation, Mean denotes the density at the mean value of the \textsc{PyMC} posterior distribution}
\end{table*}

\subsection{Analysis of the emission lines of Si IX}

The line intensities of \ion{Si}{ix} do not provide the same clear conclusions regarding the electron density and its uncertainty as \ion{Fe}{xiii}. The details of the  spectral lines used in the analysis are found in Table~\ref{tab:tab2_si_9_linelist}.

\begin{table}
	\begin{center}
	\caption{\ion{Si}{ix} line list, T = 1.12 $\times$ 10$^6$ K}
	\label{tab:tab2_si_9_linelist}
	\begin{tabular}{ccrrr}
		\hline
		$\lambda_B$ (\AA)& $\lambda_C$ (\AA & I$_{obs}$ & I$_{best}$ & I$_{pymc}$\\\hline\\
        292.806 &    292.809 &    29.8 & 33. & 32. \\		
		 $\cdots$  & 292.854 &  $\cdots$ & $\cdots$ & $\cdots$\\
		 $\cdots$  & 292.759 &   $\cdots$ & $\cdots$ & $\cdots$\\
        296.110 &    296.113 &   34.7 & 40. & 43. \\
        341.982 &    341.950 &    14.1 & 13. & 12. \\
        344.974 &    344.954 &   7.35 &  7.4 & 6.8 \\
        345.143 &    345.120 &   30.2 & 28. & 27. \\
        349.885 &    349.860 &    46.8 & 38. & 42. \\
        \hline
	\end{tabular}
	\end{center}
\raggedright{The observed line at 292.806 \AA\ is a blend of the 3 \ion{Si}{ix} lines listed. See Table~\ref{tab:tab2_fe_13_linelist} for definitions of the columns}
\end{table}

\begin{figure}
	\includegraphics[width=\columnwidth]{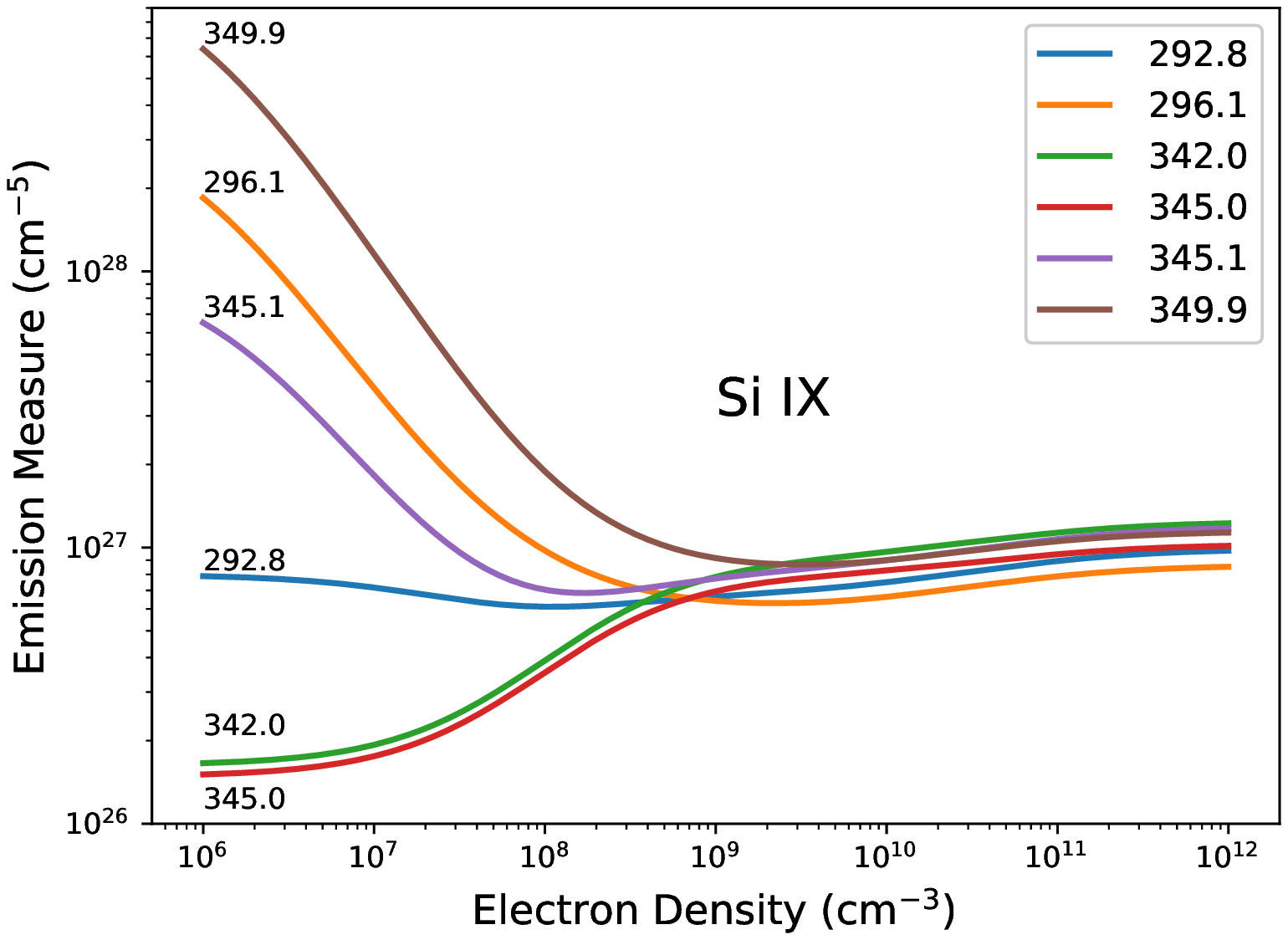}
    \caption{The simplified L-functions for the \ion{Si}{ix} lines.}
    \label{fig:si_9_emplot}
\end{figure}

The simplified L-functions for \ion{Si}{ix} are shown in Fig.~\ref{fig:si_9_emplot}. This plot suggests that the atomic model provides a good prediction of the intensities at densities above about 10$^9$ cm$^{-3}$ but that it will only be possible to derive a lower limit on the electron density.

\begin{figure}
	\includegraphics[width=\columnwidth]{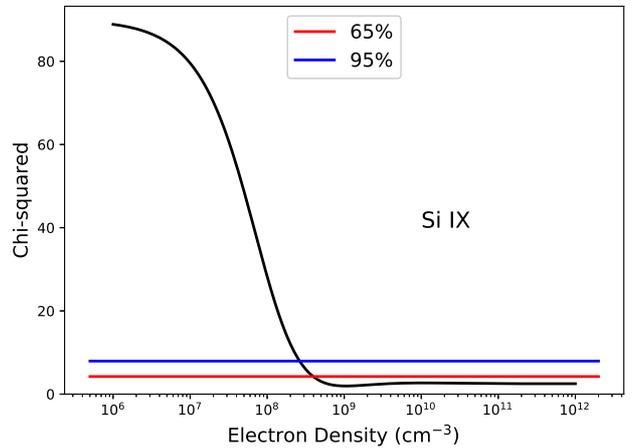}
    \caption{$\chi^2$ for the \ion{Si}{ix} lines together with the levels of 68 per cent and 95 per cent confidence.}
    \label{fig:tab2_si_9_chisquared}
\end{figure}

\begin{figure}
	\includegraphics[width=\columnwidth]{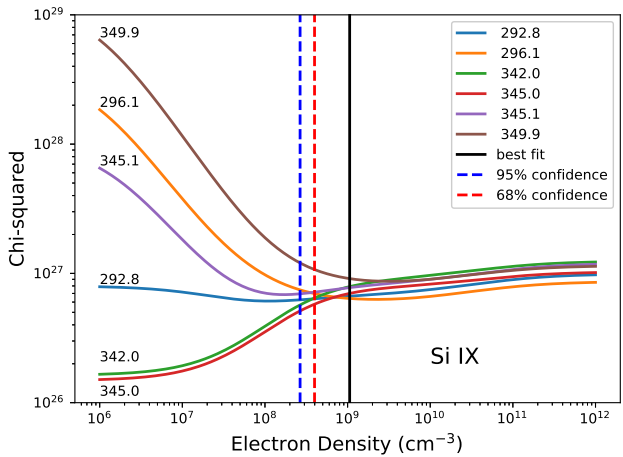}
    \caption{The simplified L-functions for the \ion{Si}{ix} lines together with the best-fitting density and the low density 95 per cent and 68 per cent confidence limits.}
    \label{fig:tab2_si_9_emplot_w_best}
\end{figure}

The $\chi^2$ minimization search produces the values of $\chi^2$ plotted in Fig.~\ref{fig:tab2_si_9_chisquared}. This plot indicates that a minimum for $\chi^2$ does exist as well as low density confidence levels. The observed intensities together with the atomic model do not establish any confidence levels at high density. The best-fitting and the lower density confidence limits are plotted in Fig.~\ref{fig:tab2_si_9_emplot_w_best}.

As suggested from the \ion{Si}{ix} simplified L-functions, the MCMC sampling does not return a very meaningful result except perhaps the lower limit to the region of 95 per cent high probability density. The MCMC posterior distribution is not very useful and not shown.

A summary of the numerical results for the $\chi^2$ minimization and the MCMC sampling are found in Table~\ref{tab:tab2_si_9_summary}.

\begin{table*}
	\begin{center}
	\caption{Summary of the analyses of \ion{Si}{ix}}
	\label{tab:tab2_si_9_summary}
	\begin{tabular}{cccccc} 
		\hline
		  95\% Conf  & 68\% Conf  & Best-fit  & 68\% Conf  & 95\% Conf & RelDev(Int)\\
		\hline
       7.1e+07 &    1.8e+08 &    4.5e+08 &    1.2e+09 &    4.0e+09  & 0.10 \\
        \hline
		    95\% HPD  &  Std  &       Mean  &       Std  &   95\% HPD & RelDev(Int)\\
    \hline
       6.3e+08 &    1.2e+09 &    1.3e+10 &    1.3e+11 &    $\cdots$ & 0.12  \\
   \hline
	\end{tabular}
	\end{center}
	\raggedright{See Table~\ref{tab:tab2_fe_13_summary} for the definitions of the entries}
\end{table*}

\subsection{Analysis of the emission lines of Mg VIII}
\label{subsec:mg8}

\begin{figure}
	\includegraphics[width=\columnwidth]{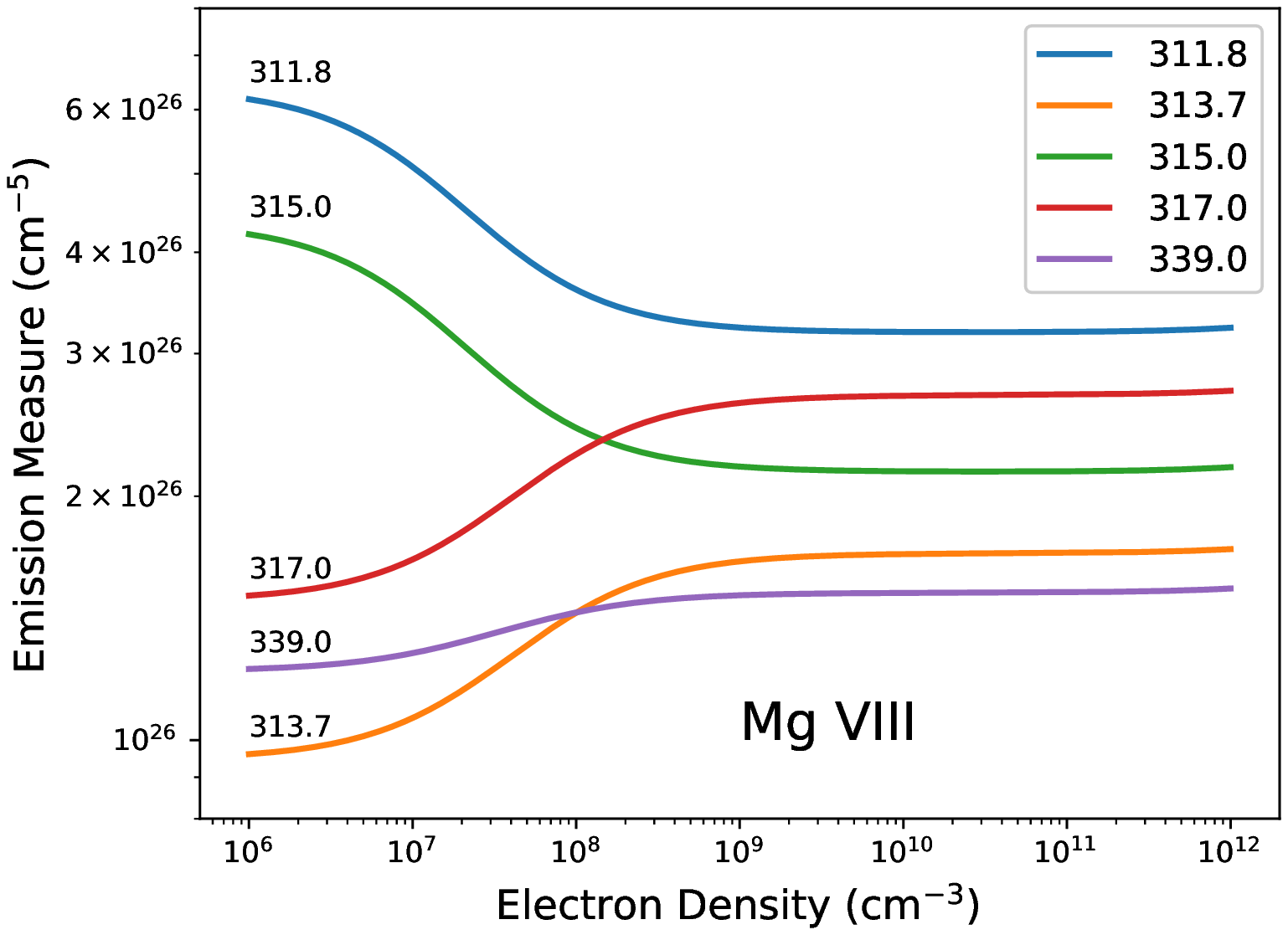}
    \caption{The simplified L-functions for the \ion{Mg}{viii} lines.}
    \label{fig:mg_8_emplot}
\end{figure}

The transitions of the \ion{Mg}{viii} spectral lines used are found in Table~\ref{tab:tab2_mg_8_linelist}. The simplified L-functions shown in Fig.~\ref{fig:mg_8_emplot} display a behaviour similar to that of the \ion{Si}{ix} L-functions except that the agreement between the intensities and the atomic model is not very good. The simplified L-function plot suggest that it will only be possible to establish some best-fitting/mean value and a lower limit to the electron density. A summary of the numerical results for the $\chi^2$ minimization and the MCMC sampling are found in Table~\ref{tab:tab2_mg_8_summary}. The values returned by the two methods are not in very good agreement.

\begin{table}
	\begin{center}
	\caption{\ion{Mg}{viii} line list, T = 7.9 $\times$ 10$^5$ K}
	\label{tab:tab2_mg_8_linelist}
	\begin{tabular}{ccrrr} 
		\hline
		$\lambda_B$ (\AA)& $\lambda_C$ (\AA)  & I$_{obs}$ & I$_{best}$ & I$_{pymc}$\\
		\hline
        311.783 &  311.772 &   13.1 &  7.9 & 7.8 \\
        313.732 &  313.743 &   12.6 &   14.4 & 14.2 \\  
        315.022 &  315.015 &   43.9 &  40. & 44. \\
        317.026 &  317.028 &   12.7 &  9.2 & 9.1\\
        339.017 &  338.983 &   6.88 &  8.8 & 8.6 \\
    \hline
	\end{tabular}
	\end{center}
\raggedright{See Table~\ref{tab:tab2_fe_13_linelist} for definitions}
\end{table}

\begin{table*}
	\begin{center}
	\caption{Summary of the analyses of \ion{Mg}{viii}}
	\label{tab:tab2_mg_8_summary}
	\begin{tabular}{cccccc} 
		\hline
		  95\% Conf  & 68\% Conf  & Best-fit  & 68\% Conf  & 95\% Conf & RelDev(Int)\\
		\hline
       6.3e+07 &    1.8e+08 &    1.2e+11 &    $\cdots$ &    $\cdots$ & 0.24 \\
        \hline
		    95\% HPD  &  Std  &       Mean  &       Std  &   95\% HPD & RelDev(Int)\\
    \hline
       1.9e+08 &    1.4e+09 &    1.6e+10 &    1.8e+11 &    $\cdots$ & 0.24 \\
   \hline
	\end{tabular}
	\end{center}
	\raggedright{See Table~\ref{tab:tab2_fe_13_summary} for the definitions of the entries}
\end{table*}

\subsection{Analysis of the emission lines of Si X}

\begin{figure}
	\includegraphics[width=\columnwidth]{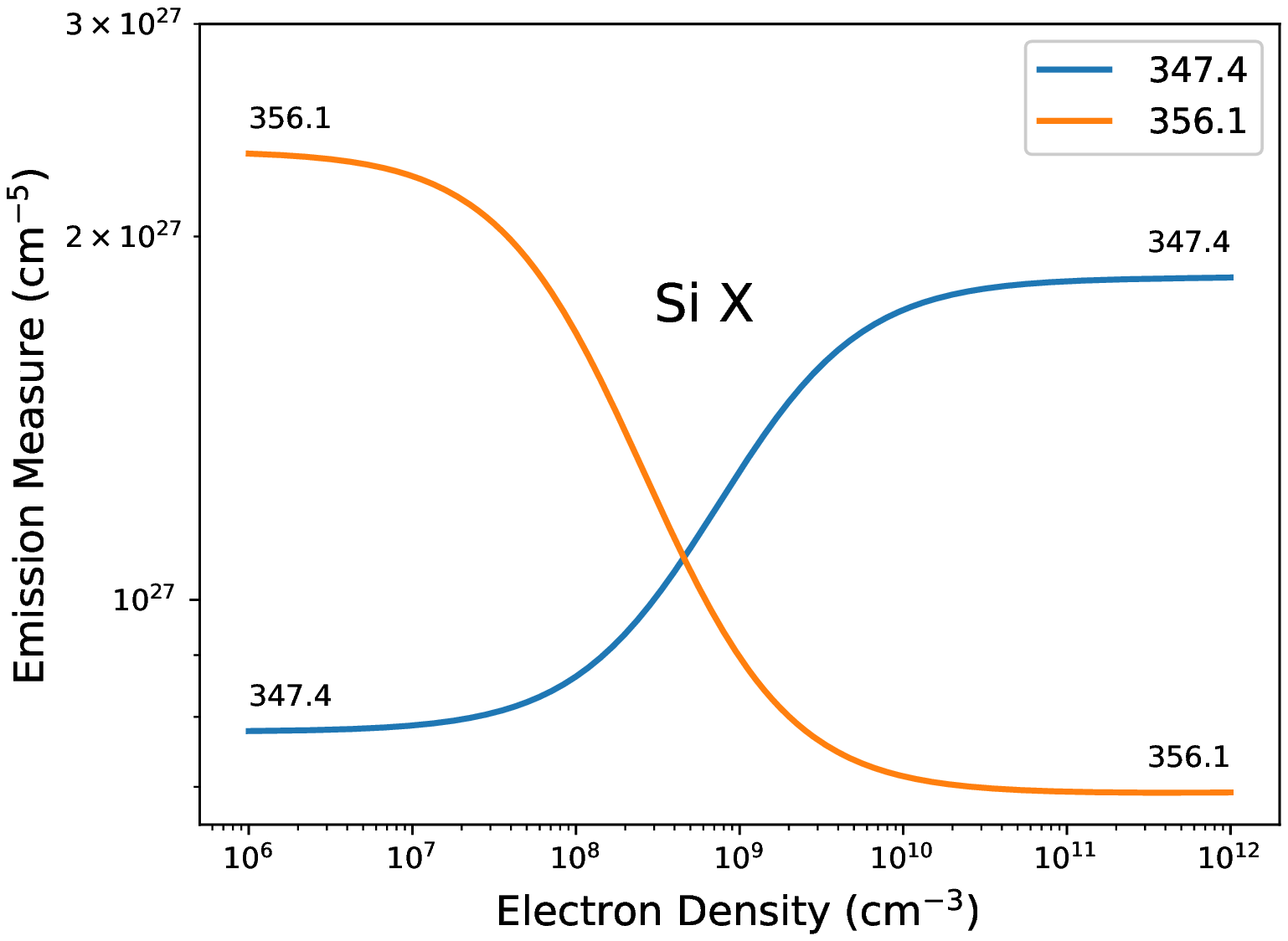}
    \caption{The simplified L-functions for the \ion{Si}{x} lines.}
    \label{fig:si_10_emplot}
\end{figure}

The details of the  spectral lines used in the analysis are found in Table~\ref{tab:tab2_si_10_linelist}. The simplified L-functions for \ion{Si}{x} are shown in Fig.~\ref{fig:si_10_emplot}. There are only two spectral lines of \ion{Si}{x} available so that one can determine the electron density from a direct comparison of the measured intensity ratio of the two lines with the predicted ratio yielding a density of 5 $\times$ 10$^8$ cm$^{-3}$.

\begin{table}
    \begin{center}
	\caption{\ion{Si}{x} line list, T = 1.41 $\times$ 10$^6$ K}
	\label{tab:tab2_si_10_linelist}
	\begin{tabular}{ccrrr}
		\hline
		$\lambda_B$ (\AA) & $\lambda_C$ (\AA)  & I$_{obs}$ & I$_{best}$ & I$_{pymc}$\\\hline
        347.419 &    347.402 &   80.9 &  81. & 73. \\		
        356.054 &    356.049 &     54.0 & 54. &   52. \\
        $\cdots$ &   356.037 &  &  &  \\
        \hline
	\end{tabular}
	\end{center}
	
See Table~\ref{tab:tab2_fe_13_linelist} for definitions
\end{table}

The results of the $\chi^2$ minimization and the MCMC modeling are found in Table~\ref{tab:tab2_si_10_summary}. The two methods yield very similar results for this fairly simple case.

\begin{table*}
	\begin{center}
	\caption{Summary of the analyses of \ion{Si}{x}}
	\label{tab:tab2_si_10_summary}
	\begin{tabular}{cccccc} 
		\hline
		  95\% Conf  & 68\% Conf  & Best-fit  & 68\% Conf  & 95\% Conf  & RelDev(Int)\\
		\hline
       7.1e+07 &    1.8e+08 &    4.5e+08 &    1.2e+09 &    4.0e+09  & 0.005 \\
        \hline
		    95\% HPD  &  Std  &       Mean  &       Std  &   95\% HPD  & RelDev(Int)\\
    \hline
       4.7e+07 &    1.3e+08 &    5.3e+08 &    2.2e+09 &    4.5e+10  & 0.06 \\
   \hline
	\end{tabular}
	\end{center}
	\raggedright{See Table~\ref{tab:tab2_fe_13_summary} for the definitions of the entries}
\end{table*}

\subsection{Analysis of the emission lines of Fe XI}
\label{subsec:fe_11}

\begin{figure}
	\includegraphics[width=\columnwidth]{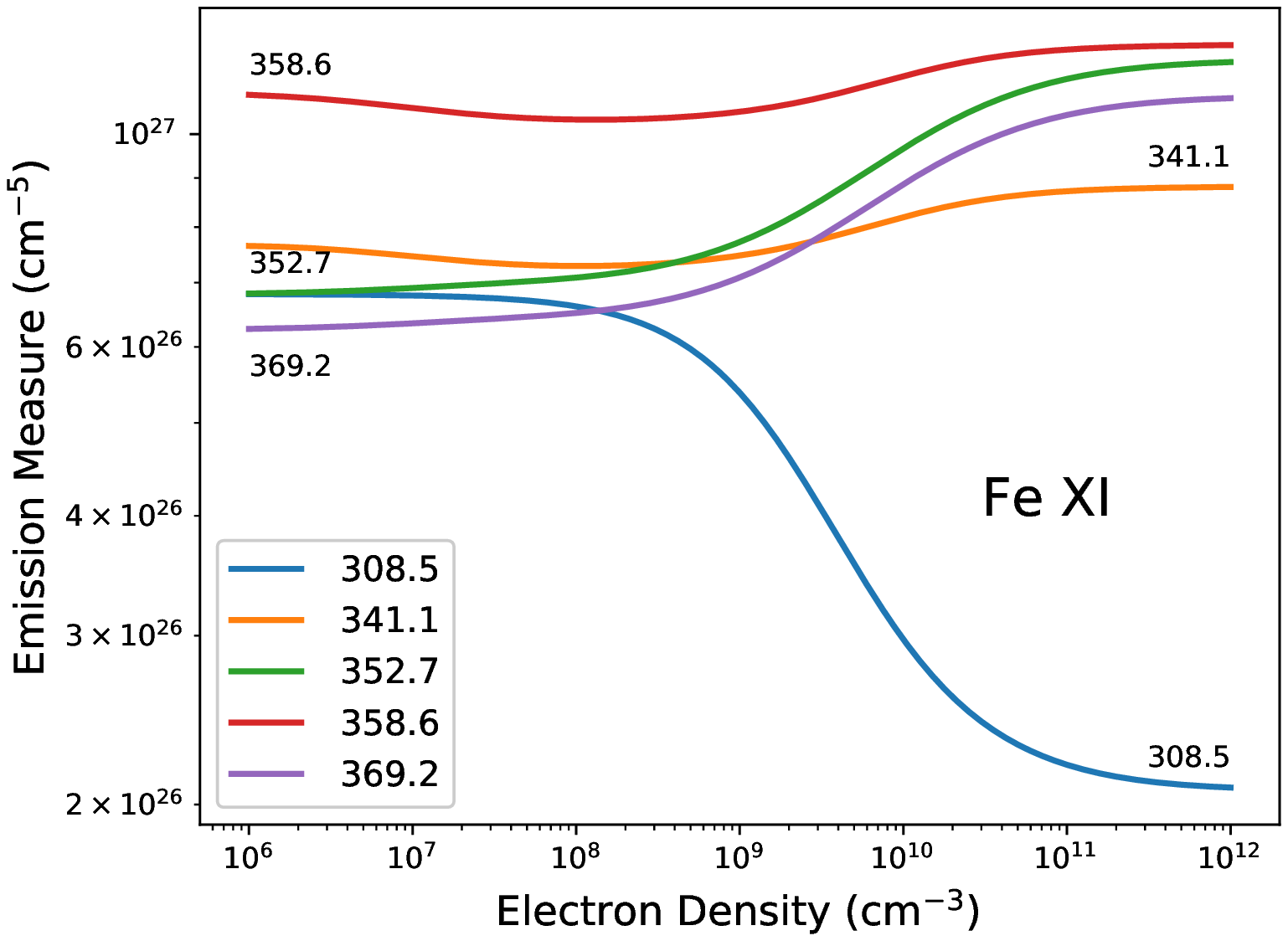}
    \caption{The simplified L-functions for the \ion{Fe}{xi} lines.}
    \label{fig:fe_11_emplot}
\end{figure}

The details of the  spectral lines used in the analysis are found in Table~\ref{tab:tab2_fe_11_linelist}. The simplified L-functions for \ion{Fe}{xi}, shown in Fig.~\ref{fig:fe_11_emplot}, reveal a situation that is reversed from that of \ion{Si}{ix}, in that one can only expect to determine an upper limit to the region of confidence. The results from the $\chi^2$ minimization and the MCMC sampling are  found in Table~\ref{tab:tab2_fe_11_summary}. The mean density from the MCMC sampling is about a factor of 3 greater than that from the $\chi^2$ minimization and the 95 per cent confidence/high-probability-density limits are with a factor of 2. The main conclusion from this analysis is that a upper limit to the 95 per cent region of confidence/high-probability-density is about 10$^9$ cm$^{-3}$. Nevertheless, the agreement between the values derived by the two methods is not very good.

\begin{table}
	\begin{center}
	\caption{\ion{Fe}{xi} line list, T = 1.26 $\times$ 10$^6$ K}
	\label{tab:tab2_fe_11_linelist}
	\begin{tabular}{ccrrr} 
		\hline
		$\lambda_B$ (\AA)& $\lambda_C$ (\AA)  & I$_{obs}$ & I$_{best}$ & I$_{pymc}$\\\hline\\
        308.533 &    308.544 &   6.46 &  7.1  & 6.9 \\		
        341.141 &    341.113 &    13.4 & 12. &  12. \\
        352.694 &    352.670 &     48.1 & 50. &  50. \\
        358.662 &    358.613 &      10.7 & 7.6 &  7.4  \\
        369.158 &    369.163 &     13.2 & 15. &  15. \\
        \hline
	\end{tabular}
	\end{center}
\raggedright{See Table~\ref{tab:tab2_fe_13_linelist} for definitions}
\end{table}

\begin{table*}
	\begin{center}
	\caption{Summary of the analyses of \ion{Fe}{xi}}
	\label{tab:tab2_fe_11_summary}
	\begin{tabular}{cccccc} 
		\hline
		  95\% Conf  & 68\% Conf  & Best-fit  & 68\% Conf  & 95\% Conf & RelDev(Int)\\
		\hline
       $\cdots$ &    $\cdots$ &    1.1e+08 &    7.5e+08 &    1.4e+09  & 0.13 \\
        \hline
		    95\% HPD  &  Std  &       Mean  &       Std  &   95\% HPD & RelDev(Int)\\
    \hline
       $\cdots$ &    6.3e+06 &    3.7e+07 & 2.2e+08 &    7.9e+08 & 0.13 \\
   \hline
	\end{tabular}
	\end{center}
	\raggedright{See Table~\ref{tab:tab2_fe_13_summary} for the definitions of the entries}
\end{table*}

\subsection{Analysis of the emission lines of Fe XII}
\label{subsec:fe_12}

\begin{figure}
	\includegraphics[width=\columnwidth]{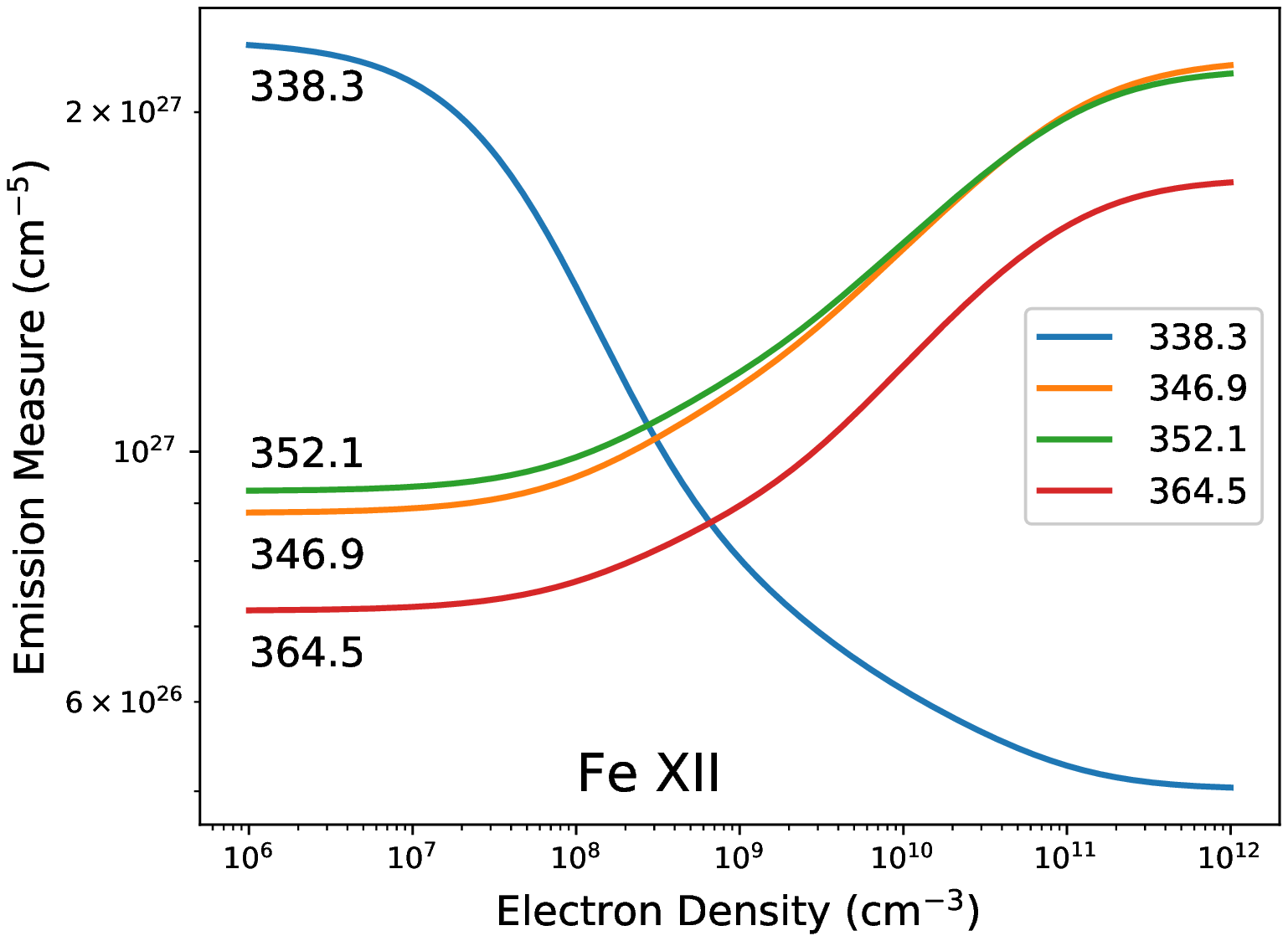}
    \caption{The simplified L-functions for the \ion{Fe}{xii} lines.}
    \label{fig:fe_12_emplot}
\end{figure}

The details of the  spectral lines used in the analysis are found in Table~\ref{tab:tab2_fe_12_linelist}. The simplified L-functions for \ion{Fe}{xii} are shown in Fig.~\ref{fig:fe_12_emplot}.  Four lines of \ion{Fe}{xii} are available for analysis. Three of the lines increase, in a relative sense, with density and only the 338 \AA\ line decreases with density. The L-functions for lines at 346 and 352 \AA\ are in good agreement with the L-function of the 364 \AA\ line lying about 20 per cent below these.

The results of the $\chi^2$ minimization and the MCMC runs are found in Table~\ref{tab:tab2_fe_12_summary}. The two methods both find a best-fitting/mean density of 4 $\times$ 10$^8$ cm$^{-3}$ and the regions of confidence from the $\chi^2$ minimization and the MCMC sampling are in good agreement.

\begin{table}
	\begin{center}
	\caption{\ion{Fe}{xii} line list, T = 1.50 $\times$ 10$^6$ K}
	\label{tab:tab2_fe_12_linelist}
	\begin{tabular}{ccccrrr} 
		\hline
		$\lambda_B$ (\AA)& $\lambda_C$ (\AA)&  I$_{obs}$ & I$_{best}$ & I$_{pymc}$\\\hline\\
        338.289 &   338.263 &   13.6 & 13.5 &  13.4 \\		
        346.871 &  346.852 &  22.3 &  20.5 &  20.3 \\
        352.127 &  352.106 &  45.7 &  41. &   40. \\
        364.490 &  364.467 &  60.3 &  70. &   70. \\
        \hline
	\end{tabular}
	\end{center}
\raggedright{See Table~\ref{tab:tab2_fe_13_linelist} for definitions}
\end{table}

\begin{table*}
	\begin{center}
	\caption{Summary of the analyses of \ion{Fe}{xii}}
	\label{tab:tab2_fe_12_summary}
	\begin{tabular}{cccccc} 
		\hline
		  95\% Conf  & 68\% Conf  & Best-fit  & 68\% Conf  & 95\% Conf & RelDev(Int) \\
		\hline
       1.6e+08 &    2.4e+08 &    4.0e+08 &    7.1e+08 &    1.0e+09  & 0.09\\
        \hline
		    95\% HPD  &  Std  &       Mean  &       Std  &   95\% HPD & RelDev(Int)\\
    \hline
       1.9e+08 &    2.8e+08 &    4.0e+08 &    5.6e+08 &    8.4e+08 & 0.09\\
   \hline
	\end{tabular}
	\end{center}
	\raggedright{See Table~\ref{tab:tab2_fe_13_summary} for the definitions of the entries}
\end{table*}

\subsection{Analysis of the emission lines of Fe XIV}
\label{subsec:fe14}

\begin{figure}
	\includegraphics[width=\columnwidth]{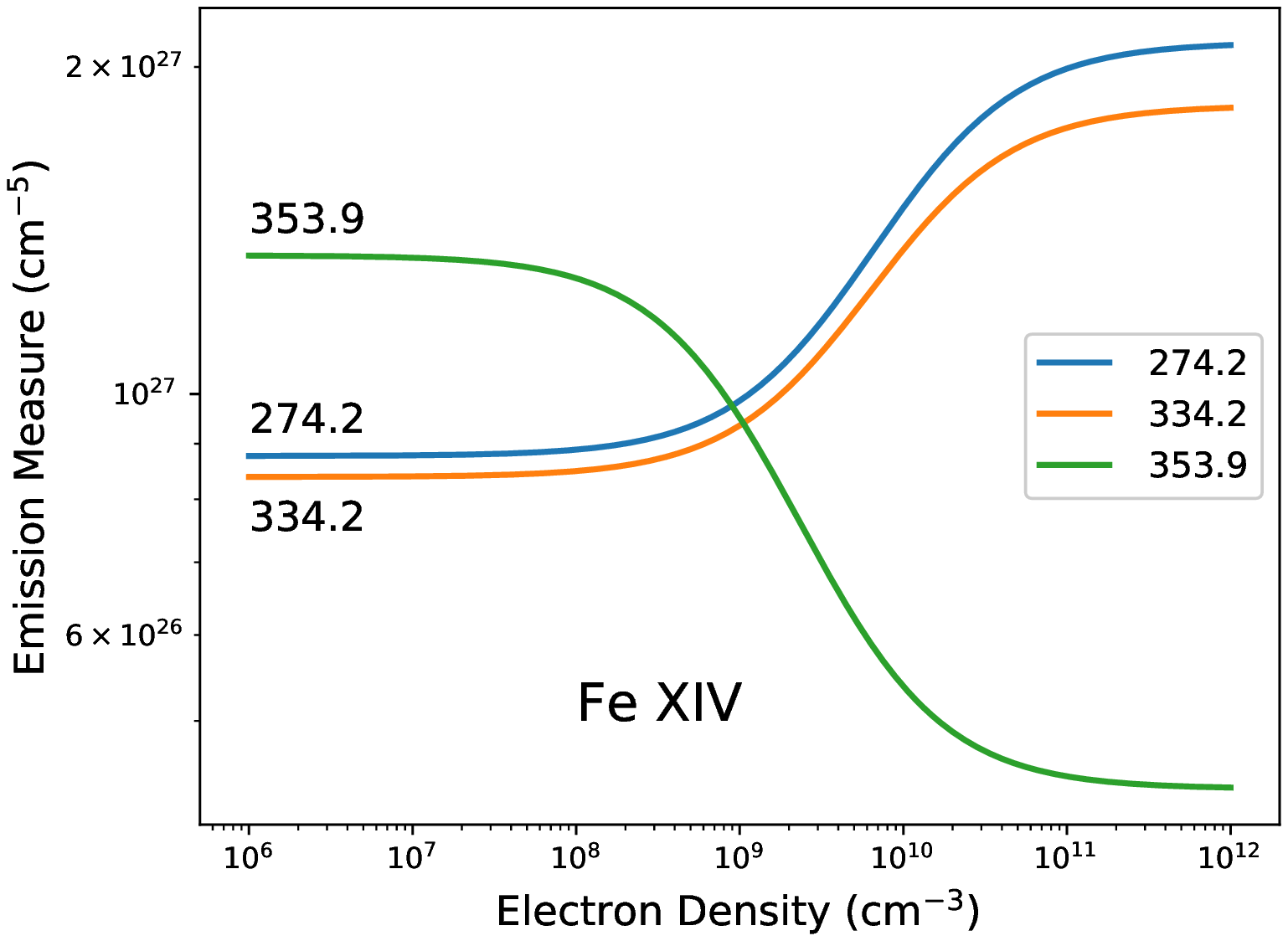}
    \caption{The simplified L-functions for the \ion{Fe}{xiv} lines.}
    \label{fig:fe_14_emplot}
\end{figure}

The details of the  spectral lines used in the analysis are found in Table~\ref{tab:tab2_fe_14_linelist}. The simplified L-functions for \ion{Fe}{xiv} are shown in Fig.~\ref{fig:fe_14_emplot}.  The L-functions for the lines at 274 and 334 \AA\ increase relatively with density and are in very good agreement and the line at 353 \AA\ decreases with density, suggesting that the errors in deriving the electron density will be relatively small.

\begin{table}
	\begin{center}
	\caption{\ion{Fe}{xiv} line list, T = 2.0 $\times$ 10$^6$ K}
	\label{tab:tab2_fe_14_linelist}
	\begin{tabular}{ccccrrr}
		\hline
		$\lambda_B$ (\AA)& $\lambda_C$ (\AA)  & I$_{obs}$ & I$_{best}$ & I$_{pymc}$\\\hline\\
        274.157 &   274.203 &   117. & 114. &  116. \\		
        334.188 &  334.178 &  79.1 &  81. &   83. \\
        353.851 &  353.836 &  26.4 &  26. &  24. \\
        \hline
	\end{tabular}
	\end{center}
\raggedright{See Table~\ref{tab:tab2_fe_13_linelist} for definitions}
\end{table}

The results of the $\chi^2$ minimization and the MCMC sampling are found in Table~\ref{tab:tab2_fe_14_summary}. The value for the best-fitting density is about 30\% higher than that derived from the mean of the MCMC posterior distribution. The upper regions of statistical confidence/high-probability-density are in good agreement while those for the lower regions are not in such good agreement. This would have been expected from an inspection of the simplified L-functions.

\begin{table*}
	\begin{center}
	\caption{Summary of the analyses of \ion{Fe}{xiv}}
	\label{tab:tab2_fe_14_summary}
	\begin{tabular}{cccccc} 
		\hline
		  95\% Conf  & 68\% Conf  & Best-fit  & 68\% Conf  & 95\% Conf & RelDev(Int)\\
		\hline
       2.1e+08 &    4.7e+08 &    9.4e+08 &    1.5e+09 &    2.0e+09 & .019 \\
        \hline
		    95\% HPD  &  Std  &       Mean  &       Std  &   95\% HPD & RelDev(Int)\\
    \hline
       8.4e+07 &    3.2e+08 &    7.1e+08 &    1.6e+09 &    1.7e+09 & .042 \\
   \hline
	\end{tabular}
	\end{center}
	\raggedright{See Table~\ref{tab:tab2_fe_13_summary} for the definitions of the entries}
\end{table*}

\section{Summary of the analysis}
\label{sec:analysis_summary}

\begin{figure}
	\includegraphics[width=\columnwidth]{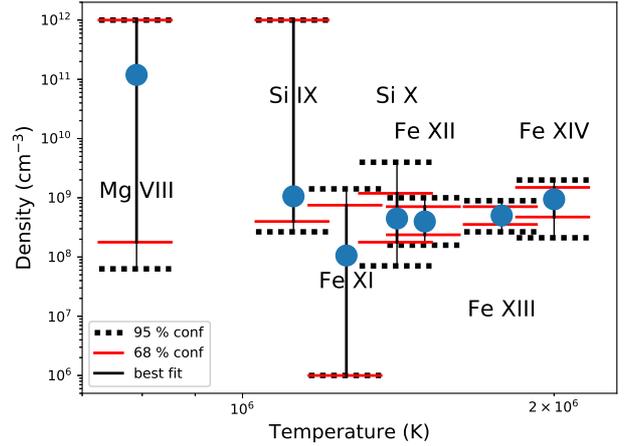}
    \caption{A comparison of the electron densities and their uncertainties, for all of the ions included, derived by means of a $\chi^2$ minimization.}
    \label{fig:tab2_br_summary}
\end{figure}

Table~\ref{tab:tab2_br_summary} summarizes all of the electron densities derived by the $\chi^2$ minimization procedure, including the best-fitting and the upper and lower limits of the regions of 68 per cent  and 95 per cent confidence. A visual comparison of the $\chi^2$ minimization analyses is shown in Fig.~\ref{fig:tab2_br_summary}. 

\begin{table*}
	\begin{center}
	\caption{Summary of the analysis performed by $\chi^2$ minimization}
	\label{tab:tab2_br_summary}
	\begin{tabular}{lcccccc} 
		\hline
		 Ion  &   T (K)      & 95\% Conf  & 68\% Conf  & Best-fit  & 68\% Conf  & 95\% Conf \\
		\hline
   Mg VIII &   7.90e+05 &    6.3e+07 &    1.8e+08 &    1.2e+11 &    \... &    1.0e+12  \\
     Si IX &   1.12e+06 &    2.7e+08 &    4.0e+08 &    1.1e+09 &    \... &    1.0e+12 \\
      Si X &   1.41e+06 &    7.1e+07 &    1.8e+08 &    4.5e+08 &    1.2e+09 &    4.0e+09 \\
     Fe XI &   1.26e+06 &    \...   &    1.0e+06 &    1.1e+08 &    7.5e+08 &    1.4e+09 \\
    Fe XII &   1.50e+06 &    1.6e+08 &    2.4e+08 &    4.0e+08 &    7.1e+08 &    1.0e+09 \\
   Fe XIII &   1.78e+06 &    2.7e+08 &    3.5e+08 &    5.0e+08 &    7.1e+08 &    8.9e+08 \\
    Fe XIV &   2.00e+06 &    2.1e+08 &    4.7e+08 &    9.4e+08 &    1.5e+09 &    2.0e+09 \\
        \hline
	\end{tabular}
	\end{center}
\raggedright{See Table~\ref{tab:tab2_fe_13_linelist} for definitions}
\end{table*}

\begin{figure}
	\includegraphics[width=\columnwidth]{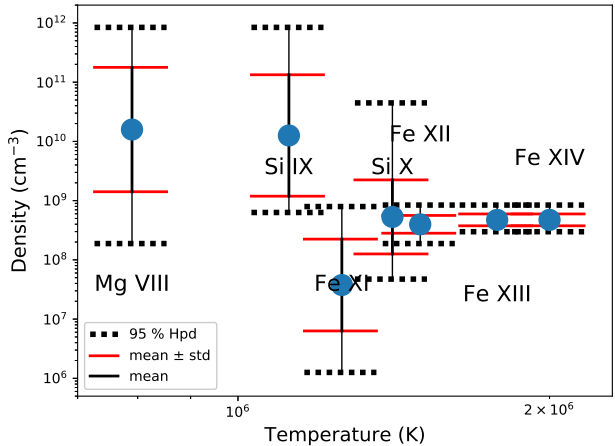}
    \caption{A comparison of the  electron densities and their uncertainties, for all of the ions included, derived by means of MCMC sampling.}
    \label{fig:tab2_mc_summary}
\end{figure}

Table~\ref{tab:tab2_mc_summary} summarizes all of the electron densities derived by the MCMC procedure, including the mean, the mean $\pm$ the standard deviation and the upper and lower limits of the regions of 95 per cent high probability density. A visual comparison of the  MCMC analyses is shown in Fig.~\ref{fig:tab2_mc_summary}. 

For both the $\chi^2$ minimization and MCMC calculations, the best-fitting/mean densities for \ion{Fe}{xii, xiii, xiv} and \ion{Si}{x} are about 5 $\times$ 10$^{8}$ cm$^{-3}$. I will refer to these ions as the {\it robust} ions. The ions \ion{Mg}{viii} and \ion{Si}{ix} provide lower limits that are roughly consistent with the the lower limit suggested by the {\it robust} ions. The ion \ion{Fe}{xi} provides an upper limit that is consistent with the upper limits provided by the {\it robust} ions.

The fact that the electron densities derived from the {\it robust} ions are all roughly the same supports the idea that the measurements are an average over a number of features. For example, \citet{dere_eis_qs} analyzed quiet Sun spectra of \ion{Fe}{xii} and \ion{Fe}{xiii} obtained by the Extreme Ultraviolet Imaging Spectrometer (EIS) instrument on {\it Hinode}. They found densities ranging from 3 to 20 $\times$ 10$^8$ cm$^{-3}$, with the highest values found in bright-points

\begin{table*}
	\begin{center}
	\caption{Summary of the analysis performed by MCMC}
	\label{tab:tab2_mc_summary}
	\begin{tabular}{lcccccc} 
		\hline
		 Ion  &   T      &    95\% HPD  &  Std  &       Mean  &       Std  &   95\% HPD \\
		\hline
   Mg VIII &   7.90e+05 &    1.9e+08 &    1.4e+09 &    1.6e+10 &    1.8e+11 &    8.4e+11 \\
     Si IX &   1.12e+06 &    6.3e+08 &    1.2e+09 &    1.3e+10 &    1.3e+11 &    8.4e+11 \\
      Si X &   1.41e+06 &    4.7e+07 &    1.3e+08 &    5.3e+08 &    2.2e+09 &    4.5e+10 \\
     Fe XI &   1.26e+06 &    1.3e+06 &    6.3e+06 &    3.8e+07 &    2.2e+08 &    7.9e+08 \\
    Fe XII &   1.50e+06 &    1.9e+08 &    2.8e+08 &    4.0e+08 &    5.6e+08 &    8.4e+08 \\
   Fe XIII &   1.78e+06 &    3.0e+08 &    3.8e+08 &    4.7e+08 &    6.0e+08 &    8.4e+08 \\
    Fe XIV &   2.00e+06 &    8.4e+07 &    3.2e+08 &    7.1e+08 &    1.6e+09 &    1.7e+09 \\
        \hline
	\end{tabular}
	\end{center}
\raggedright{See Table~\ref{tab:tab2_fe_13_linelist} for definitions}
\end{table*}

\section{Discussion}
\label{sec:discussion}

\subsection{Si IX}
\label{subsec:discussion_si9}

The current CHIANTI atomic model is described in \citet{landi_v2}. It uses the R-matrix calculations of \citet{aggarwal_si9} to describe the transitions among the ground configuration and the DW calculations of \citet{bhatia_si9} for transitions among the higher levels.

The results for \ion{Si}{ix} are not very meaningful. The $\chi^2$ minimum is very shallow and does not really agree with the mean density derived from the MCMC sampling. \citet{brosius} did not use the \ion{Si}{ix} lines to derive electron densities. Using quiet Sun spectra from the Coronal Diagnostics Spectrometer (CDS) experiment on {\it SOHO}, \citet{landi_nis_qs_ar} derived an electron density of 5 $\times$ 10$^8$ cm$^{-3}$. Using spectra from the same instrument,  \citet{young_qs_fip} found that the electron densities derived from \ion{Si}{ix} were fairly constant at a value of about 3 $\times$ 10$^8$ cm$^{-3}$. These latter results refer to a combination of quiet Sun and coronal hole.

\subsection{Mg VIII}
\label{subsec:discussion_mg8}

The components for constructing the atomic model are reported in \citet{delzanna_v8}. The collision rates are based on the R-matrix calculations of \citet{liang_bseq}. \citet{brosius} did not report electron densities from the \ion{Mg}{viii} lines, perhaps due to the same difficulties reported in Section~\ref{subsec:mg8}. \citet{landi-L-function} also examined the \ion{Mg}{viii} density sensitive line ratios in two quiet Sun regions using data from CDS. They show similar L-function plots to what we show in Fig.~\ref{fig:mg_8_emplot}. In the less intense region, they find only a lower limit of 10$^8$ cm$^{-3}$.

\subsection{Si X}
\label{subsec:discussion_si10}

The atomic parameters for constructing the atomic model used here are reported in \citet{delzanna_v8} and are based on the R-matrix calculations of \citet{liang_bseq}.

\citet{brosius} did not report electron densities from the \ion{Si}{x} lines. \citet{landi_nis_qs_ar} reported electron densities of 4 $\times$ 10$^8$ and 6 $\times$ 10$^8$ \cm\ in two quiet regions of the Sun, although it is not clear what spectral lines they actually used.  \citet{landi_nis_qs_ar} used atomic data based on the R-matrix calculations of \citet{zhang_bseq}. These calculations should be of comparable quality to the CHIANTI model for the transitions examined here. \citet{kamio_qs} found that the electron density determined from observations in the quiet Sun varied from about 3 $\times$ 10$^8$ to about 5 $\times$ 10$^8$ \cm\ over a 4 year period of EIS synoptic observations. It appears that these various measurements of electron densities derived from \ion{Si}{x} are in general agreement.  For comparison, the densities derived here are 4.5 $\times$ 10$^8$ for the best-fitting value and 1.3 $\times$ 10$^{10}$ for the mean of the MCMC posterior distribution.

\subsection{Fe XI}
\label{subsec:discussion_fe11}

The current version of CHIANTI includes the recent R-matrix calculations of \cite{delzanna_fe11}. \citet{brosius} determined an electron density of about 2.5 $\times$ 10$^{9}$ cm$^{-3}$, about a factor of 10 higher than derived here, 1.1 $\times$ 10$^8$ and 3.7 $\times$ 10$^7$ cm$^{-3}$. Their atomic model was based on the distorted-wave calculations of \citet{mason_fe11}.   One would expect that the distorted wave (DW) calculations would underestimate the collision rates within the ground configuration whose population largely governs the density sensitivity of the observed lines. \citet{warren_qs} analyzed a number of density sensitive line ratios in the quiet Sun with EIS spectra. In the case of \ion{Fe}{xi}, they derive an electron density of 1.4 $\times$ 10$^8$ \cm, but, considering the shallowness of the density ratio curve, the error could be considerable.

\subsection{Fe XII}
\label{subsec:discussion_fe12}

The electron densities derived by \citet{brosius} for \ion{Fe}{xii} are about a factor of 2 higher than value of 4 $\times$ 10$^8$ cm$^{-3}$ reported in Section~ \ref{subsec:fe_12}. The atomic model of \citet{brosius} used the R-matrix collision strengths for the ground configuration of \citet{tayal_fe12}. These should provide  diagnostic ratios comparable to the current CHIANTI model, described by \citet{delzanna_v8}, based on the R-matrix calculations of \citet{delzanna_fe12}. Consequently, there is no simple answer for the differences. It should be noted that \citet{brosius} only used the ratio of the 338 \AA\ line to the 352\AA\ line. If the 364 \AA\ line had been included, as in this analysis, the density they derived would be somewhat higher.

\citet{dere_eis_qs} was able to map the electron densities in the quiet Sun at a spatial resolution of about 2 arcsec with EIS. Within the quiet Sun, the darker regions had a density of about 2.5 $\times$ 10$^8$ cm$^{-3}$ and the brighter regions a density of about 5 $\times$ 10$^8$. These are in good agreement with the average density derived from SERTS. \citet{landi_nis_qs_ar} derive values of electron densities of 1.4-2.0 $\times$ 10$^9$ cm$^{-3}$ in the quiet Sun, considerably higher than found here. Their work was based on the Arcetri Spectral Code but there is little information in literature about it. The atomic data was probably similar to that used by \citet{brosius}. \citet{warren_qs} derived quiet Sun densities of about 3 $\times$ 10$^8$ \cm, closer to what is found here. They used the atomic data from CHIANTI version 5 \citep{landi_v5} that includes the R-matrix calculations of \citet{storey_fe12} that should be of comparable accuracy as those of \citet{delzanna_fe12} that are used here.

\subsection{Fe XIII}
\label{subsec:discussion_fe13}

The densities that \citet{brosius} reported for the 1993 QS observations of \ion{Fe}{XIII} ranged from 6 $\times$ 10$^8$ to  2 $\times$ 10$^9$ cm$^{-3}$, with a rough average of 1 $\times$ 10$^9 $cm$^{-3}$. This is about a factor of 2 higher than 5 $\times$ 10$^8$ cm$^{-3}$ reported here in Section~\ref{subsec:fe_13} . The collision data that \citet{brosius} used were the distorted wave calculations of \citet{fawcett_mason_fe13}. The current version 9 of the CHIANTI database contains the \citet{delzanna_fe13} R-matrix collision strengths and should provide a more accurate calculation of the ground configuration populations. \citet{landi_nis_qs_ar} derive values of electron densities of 6-14 $\times$ 10$^8$ \cm\ in the quiet Sun, comparable to \citet{brosius}. \citet{warren_qs} derived quiet Sun densities of about 2 $\times$ 10$^8$ \cm, closer to what is found here.

\subsection{Fe XIV}
\label{subsec:discussion_fe14}

The current CHIANTI version 9 database includes the collision calculations of \citet{liang_fe14} who performed an R-matrix calculation over the lowest 197 fine structure levels. The CHIANTI model is described by \citet{landi_v7}.

\citet{brosius} derived a value of the electron density of 2.2 $\times$ 10$^9$  cm$^{-3}$ from the ratio of the 353 and 335\AA\ lines. This is roughly a factor of 3 higher than the best-fitting and mean densities listed in Table~\ref{tab:tab2_fe_14_summary}, 1.5 $\times$ 10$^9$ and 1.6 $\times$ 10$^9$ cm$^{-3}$, respectively. \citet{brosius} used the collision calculations of \citet{dufton_fe14} that was an early implementation of the R-matrix package and involved the calculation of collision strengths between the 6 lowest LS states. Collision strengths between the 12 fine structure levels were determined by means of a term-coupling scheme.

One of the more important collision rates governing the population of the 3s$^2$3p $^2$P$_{3/2}$ metastable level is that from the ground level in the same configuration. At a temperature of 2 $\times$ 10$^6$ K, the collision strength from the \citet{liang_fe14} calculations is about a factor of 1.8 higher than that of \citet{dufton_fe14}. This explains some of the factor of 3 difference in the densities.

\citet{warren_qs} derived quiet Sun densities of about 2 $\times$ 10$^9$ \cm, considerably higher than the other coronal densities they measured but closer to the quiet Sun density found by \citet{landi_nis_qs_ar}.

\section{Conclusions}
\label{sec:conclusions}

Three methods have been used in the analysis of density sensitive line intensities to determine mean and best-fitting electron densities from observed spectra as well as regions of statistical confidence. The use of simplified L-functions provides a quick visual means of determining a rough estimate for these quantities. Two quantitative methods have also been employed, a $\chi^2$ minimization procedure with the criteria of \citet{lampton} and a Bayesian inference approach by means of a MCMC calculation implemented with the \textsc{Python} \textsc{PyMC} package. A $\chi^2$ minimization search together with the \citet{lampton} criteria allows one to determine the best-fitting value and the regions for a specified degree of statistical confidence. The posterior distributions of the MCMC calculations allow the determination of the mean density and the regions of high probability density to any specified degree. For ions whose L-functions indicate a robust determination of densities and errors, the agreement between the $\chi^2$ minimization and the MCMC sampling is good. For other ions, whose L-function diagrams are not so robust, only lower or upper limits on the electron density can be determined. These tend to agree with the limits derived from the more robust ions.

It should also be noted that the electron densities derived here are generally significantly lower than those derived by \citet{brosius}. The main conclusion is the newer scattering calculations with the R-matrix codes is better able to compute the collision strengths for forbidden and intercombination transitions than the distorted wave calculations that were available to \citet{brosius}.

As mentioned in Section~\ref{sec:Intro}, the use of density sensitive line intensities to determine electron densities has been going on for some time. The primary goal of this paper has been to report methods that can be successfully used to determine the best-fitting/mean values of electron density as well as their regions of statistical confidence, in the case of the $\chi^2$ minimization, or the regions of high probability density in the case of the MCMC Bayesian approach.

\section*{Acknowledgements}

I am very grateful to Dr. Jeffrey Brosius for providing the tables from \citet{brosius} in a machine readable format. The NASA Astrophysics Data System has been an invaluable resource for this project. This work as been supported by NASA grants NNX15AF25G and NNX15AF48G.




\bibliographystyle{mnras}
\bibliography{serts-densities} 




\bsp	
\label{lastpage}
\end{document}